\documentclass[prb, nobibnotes, aps, preprint, floatfix, superscriptaddress, notitlepage]{revtex4-2}
\usepackage[T1]{fontenc}
\usepackage[utf8]{inputenc}
\usepackage{amsmath,amssymb}
\usepackage[margin=2cm]{geometry}
\usepackage{graphicx}

\usepackage[colorlinks, allcolors=blue]{hyperref}
	\hypersetup{linktocpage}

\usepackage{amsmath}
\usepackage{amssymb}

\begin{document}

\title{
        Tellurite glass rods with nanodiamonds as photonic magnetic field and temperature sensors
}

\author{Zuzanna Orzechowska}
\affiliation{M. Smoluchowski Institute of Physics, Jagiellonian University in Krak\'ow, 30-348 Krak\'ow, Poland}

\author{Mariusz Mr\'ozek}
\affiliation{M. Smoluchowski Institute of Physics, Jagiellonian University in Krak\'ow, 30-348 Krak\'ow, Poland}

\author{Adam Filipkowski}
\affiliation{Department of Photonics, University of Warsaw, 02-093 Warsaw, Poland}

\author{Dariusz Pysz}
\affiliation{Łukasiewicz Research Network, Institute of Microelectronics and Photonics, 02-668 Warsaw, Poland}

\author{Ryszard St\k{e}pie\'n}
\affiliation{Łukasiewicz Research Network, Institute of Microelectronics and Photonics, 02-668 Warsaw, Poland}

\author{Mateusz Ficek}
\affiliation{Department of Metrology and Optoelectronics, Gdańsk University of Technology, Gda\'nsk 80-233, Poland}

\author{Adam M. Wojciechowski}
\email{a.wojciechowski@uj.edu.pl}
\affiliation{M. Smoluchowski Institute of Physics, Jagiellonian University in Krak\'ow, 30-348 Krak\'ow, Poland}

\author{Mariusz Klimczak}
\affiliation{Department of Photonics, University of Warsaw, 02-093 Warsaw, Poland}
\affiliation{Łukasiewicz Research Network, Institute of Microelectronics and Photonics, 02-668 Warsaw, Poland}

\author{Robert Bogdanowicz}
\affiliation{Department of Metrology and Optoelectronics, Gdańsk University of Technology, Gda\'nsk 80-233, Poland}

\author{Wojciech Gawlik}
\affiliation{M. Smoluchowski Institute of Physics, Jagiellonian University in Krak\'ow, 30-348 Krak\'ow, Poland}


\begin{abstract}
We present the results of work on a hybrid material composed of a tellurite glass rod doped with nanodiamonds containing nitrogen-vacancy-nitrogen and paramagnetic nitrogen-vacancy color centers. The reported results include details on tellurite glass and cane fabrication, confocal and wide-field imaging of the nanodiamond distribution in their volume, as well as on the spectroscopic characterization of their fluorescence and Optically Detected Magnetic Resonance measurements of magnetic fields and temperatures. Magnetic fields up to 50 G were examined with a sensitivity of $10^{-5}$ T Hz$^{-1/2}$ whereas temperature measurements were simultaneously performed with a sensitivity of 74 kHz K$^{-1}$ within the 8 Kelvin range at room temperature. In that way, we demonstrate the suitability of such systems for fiber magneto- and thermometry with a reasonable performance already in the form of glass rods. At the same time, the rods constitute an interesting starting point for further processing into photonic components such as microstructured fibers or fiber tapers for the realization of specialized sensing modalities.
\end{abstract}

\maketitle{}

\section{Introduction}

Nitrogen-vacancy-nitrogen (NVN) and negatively charged nitrogen-vacancy (NV$^{-}$) color centers are point defects in the diamond lattice which find increasingly more important applications in photonics, sensors, and biomedical diagnostics.

The NVN (also known as H3), structure is formed by a nitrogen pair trapped near a vacancy and is one of the most commonly present lattice defects in natural diamonds after the thermal annealing process. In particular, the H3 color centers are used as in vivo fluorescent markers due to their low toxicity, strong green fluorescence~\cite{yu2005bright,chang2008mass,smith2007transfection}, and long fluorescent lifetime~\cite{nunn2019brilliant}. Very good thermal stability, even at temperatures as high as 500$^{\circ}$C, makes NVN centers interesting candidates for room temperature color center lasers~\cite{rand1985visible}. H3 centers can be taken into consideration in thermometry applications based on the frequency shift of their zero-phonon line (ZPL)~\cite{davies1974vibronic}.

The NV$^-$ are diamond defects where a substitutional nitrogen atom replaces the carbon and an adjacent lattice site is left empty creating a vacancy. In addition to applicability as fluorescent markers for biological systems caused by their strong red fluorescence~\cite{Wojciechowski2018, Balasubramanian2008}, the NV$^-$ centers are paramagnetic, hence exhibit very interesting magnetic properties. For example, they can be used as qubits in quantum information science~\cite{Golter2016,Jamonneau2016}, and sensors of electric~\cite{Schirhagl2013} or magnetic~\cite{acosta2009diamonds,Wolf2015} fields. 

Large sizes of typical commercial high-quality magnetometers motivate new developments of small-scale devices combining optical fibres with color centers in diamonds. That approach was realised in several ways: by attaching a diamond to the optical fiber~\cite{van2009nanopositioning,fedotov2014electron,dmitriev2016concept}, incorporating it into photonic crystal fibres~\cite{schroder2011fiber}, growing the diamond material at the outer fiber core~\cite{bai2020fluorescent}, coupling them with tapered optical fibres~\cite{liu2013fiber}, or designing miniaturized hybrid magnetometers~\cite{hatano2021simultaneous,sturner2021integrated}.

One promising direction of these developments is based on the melt mixing of diamonds and glass material during the drawing process. Such an approach has been notably advanced by the Monro group~\cite{ebendorff2014nanodiamond,ruan2015nanodiamond,ruan2018magnetically}. We followed their recipe for glass preparation and similarly chose the tellurite glass which is well-suited as a host material for nanodiamonds. The reason for that is the relatively low melting point temperature of 600 to 700$^{\circ}$C minimizing the oxidation of nanodiamonds (NDs) when mixing them into the glass melt. Moreover, tellurite dioxide (TeO$_2$) based glasses are transparent in the range of 500 – 4000 nm and possess highly nonlinear optical properties~\cite{yamada2003ultra}. Embedding NDs in a tellurite glass host enables efficient coupling of single-photon emission to the bound modes of rods drawn from this material. These features, together with high transparency and low transmission losses (around 1 dBm$^{-1}$ for wavelengths 550 - 900 nm~\cite{Oermann:09}) determined mainly by the NV$^-$ centers absorption are essential in exploiting the unique properties of  NV$^{-}$ doped NDs and application of such material in various photonic devices.

In addition to their interesting optical properties, a more important feature of the negatively charged NV defect is its magnetic properties due to the nonzero electronic spin (S=1) that enables magnetic resonance studies and, together with a long coherence time, makes these centers suitable for sensing applications. The sensing principles are extensively described (e.g., in Refs.~\cite{rondin2014magnetometry, acosta2011optical, kimball2013general}), so we just briefly summarize the main features which make it possible. The ground state of the color center defect NV$^-$ is a spin triplet. In the absence of a magnetic field, the ground state sublevels m$_s$ = 0 and m$_s$ = $\pm$ 1 are split by 2.88 GHz, while the application of a magnetic field splits the m$_s$~=~$\pm$ 1 levels with a slope of 2.8 MHz per G of the field component along the NV axis. Using optical pumping (excitation by a green light), the ground state of the NV$^-$ defect becomes spin-polarized. Application of the resonant microwave (MW) field allows the manipulation of the m$_s$ sublevel populations with energy splitting depending on the magnetic field projection on the NV axis and temperature, while the fluorescence signal can be used to probe the spin-state population difference in an optically detected magnetic resonance (ODMR) measurement. The ODMR experiments performed here are based on collecting the red fluorescence from the tellurite glass rods enriched with NVs as a function of the MW frequency and magnetic field.

In this paper, we describe the development of incorporating two kinds of fluorescent NDs into tellurite glass for photonic sensing applications. We use high-pressure high-temperature (HPHT) nanodiamonds of size around 750 nm, with both nitrogen-vacancy-nitrogen and nitrogen-vacancy color centers. The former offers green fluorescence (with the maximum at $\sim$520 nm), while the latter fluorescein in the red ($\sim$670 nm), which allows for their independent detection and addressing. In the studied samples, we observe the contributions of the NVN and NV centers and extend the previous studies by demonstrating measurements of magnetic fields in a wide range, up to 50 G, and by employing the thermometry based on the ODMR. Glass rods doped with color centers appear to be a convenient material platform for further modification for the development of various photonic devices, e.g., ending of rods with needle shapes, tapers, fibre extruding, making microlenses, plates, etc.

\section{Experimental Results}

For initial characterization of the fabricated tellurite glass rods, the distribution of diamond particles was investigated with the help of confocal fluorescence microscopy, which is described in Section 2.1. This allowed us to evaluate the doping process, particularly the amount of NDs detectable in the drawn rod. In Section 2.2, we analyze the fluorescence spectra of the doped rod in order to compare them with the raw ND material. Finally,  Section 2.3. is devoted to the observation of optically detected magnetic resonance signals and its use for sensing applications in the wide-field microscopy setup. 

\subsection{Confocal Imaging}

 \begin{figure}
    \centering
    \includegraphics[width=0.80\columnwidth]{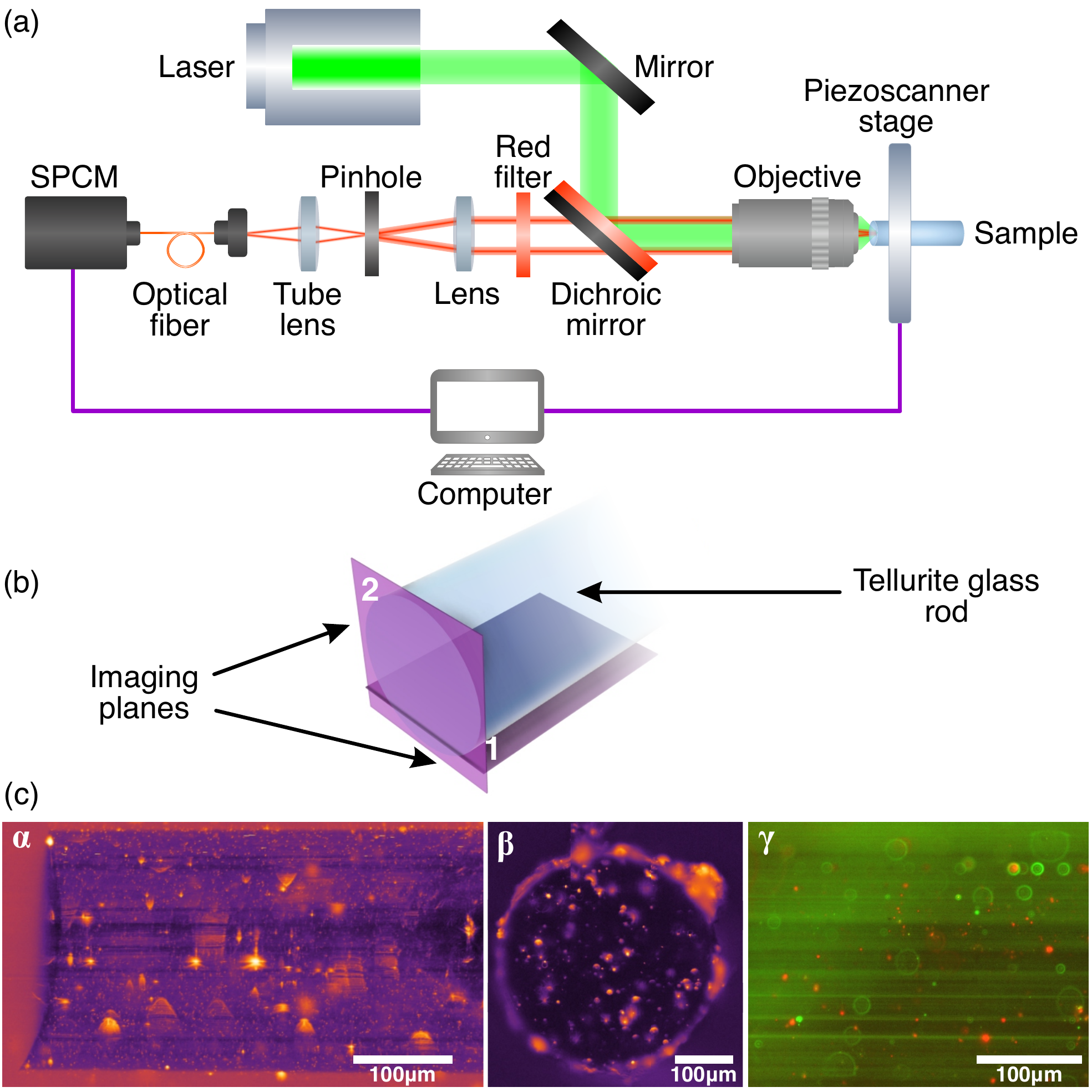}
    \caption{(a) Beam path and optical elements used in the confocal setup. (b) Schematic view of the tellurite rod with indicated planes 1 and 2 which correspond to images depicted below. (c) Images of the front face ($\alpha$) and side of the rod ($\beta$) were recorded with the scanned confocal microscope. Image $\alpha$ was collected with the rod of 245 $\mu$m diameter at the depth of 42 $\mu$m. Image $\beta$ shows the front face of the 400 $\mu$m diameter rod (bright spots on the outer edge of its face are recognized as dust particles and are also visible with a standard optical microscope). Both images $\alpha$ and $\beta$ are a result of merging 6 and 9, 200x200 $\mu$m single scans, respectively. Images $\alpha$ and ${\beta}$ were taken with red spectral filters transmitting the NV$^{-}$ emission, whereas image $\gamma$ was taken with different spectral filters selectively transmitting the green (NVNs) and red (NV$^{-}$s) fluorescence.} 
    \label{fig:confocal}
\end{figure}

Figure \ref{fig:confocal}(a) schematically presents the beam paths and relevant optical components of the home-built confocal setup used for imaging NDs containing NV centers in the rod volume. Two-dimensional sample scans were performed along the planes indicated schematically in Fig.~\ref{fig:confocal}(b). The resulting fluorescence images recorded are shown in Fig.~\ref{fig:confocal}(c). Images $\alpha$ and $\beta$, obtained for planes marked 1 and 2, respectively, visualize the NDs containing NV centers as bright spots. To visualize both kinds of NDs, i.e. also those containing NVN centers, the image $\gamma$ in Figure~\ref{fig:confocal}(c) was obtained with a commercial microscope equipped with several excitation sources and suitable filter sets. By merging the images recorded at the green and red wavelength band under blue and green excitation, respectively, we confirmed that the developed tellurite glass contains comparable densities of NDs with the NVN and NV${^-}$ color centers. 

The amount of diamonds present in the glass matrix after rod fabrication may be affected by many factors, such as melting temperature, glass cane drawing speed at the fibre drawing tower, the concentration of NDs added to the melt, and the size of nanodiamonds~\cite{ebendorff2014nanodiamond,ruan2015nanodiamond}. As mentioned in the 4.1. Sample Preparation subsection of this paper, the concentration of NDs at the input was 0.001\% (w/w) and from the obtained images we were able to calculate the average number of nanodiamonds with color centers that were visible after (survived) the drawing process. We use the same methodology as presented in Ref.~\cite{ruan2015nanodiamond}. 25 NDs with both NV$^{-}$ and NVN color centers were identified on 6 merged confocal images of known focal depth 1.44 $\mu$m and the volume of around 4.2$\times10^5$ $\mu$m${^3}$. The initial number of diamond particles incorporated into the glass was around 2.6$\times10^9$. Based on the confocal images acquired and assuming the identical distribution of NDs along the full length of the drawn rod, around 93\% of the initial number of diamond particles survived the drawing process which is significantly more than in Ref.~\cite{ruan2015nanodiamond}. This confirms our prediction that larger-sized diamonds are more resilient for mixing with glass and the tellurite glass with a low melting point is a well-suited host material for diamonds with NV$^{-}$ and NVN color centers.

\subsection{Fluorescence Spectra}

\begin{figure}
    \centering
    \includegraphics[width=0.80\columnwidth]{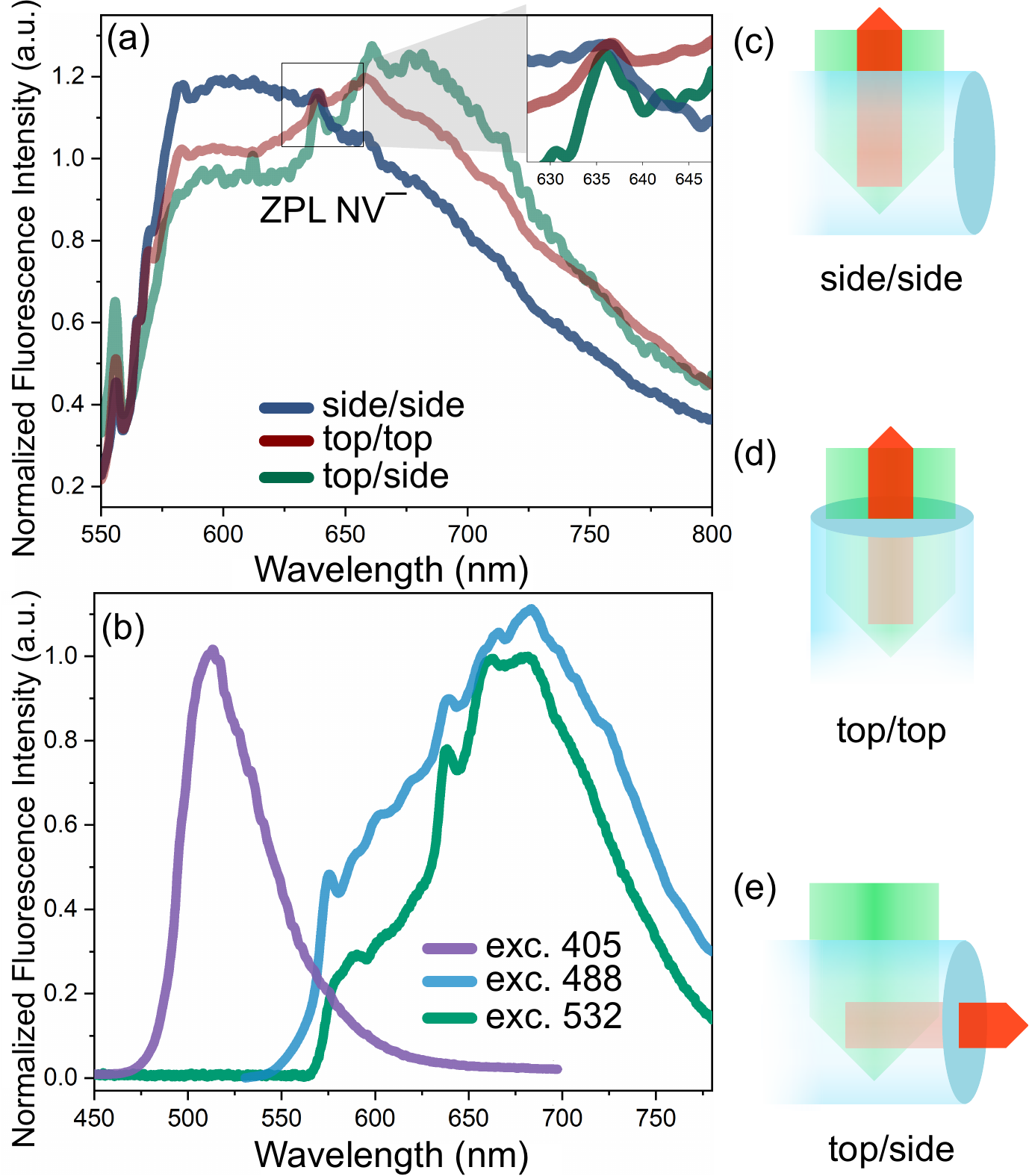}
    \caption{(a) Fluorescence spectra of the ND-doped tellurite glass rod excited at 532 nm. The three configurations of the excitation and detection directions are depicted on (c), (d), and (e), and color-labelled in the figure legend as excitation direction/detection direction. The marked area in the inset shows ZPL of NV${^-}$ at 637 nm. (c), (d), (e) shows geometrical orientations of the excitation beam (green) and fluorescence collection direction (red) in reference to the orientation of the tellurite rod (blue cylinder). (b) Fluorescence spectra of the starting material MDNVN1 powder were obtained for excitation wavelengths 405 nm, 488 nm, and 532 nm with a proper set of filters.} 
    \label{fig:spectrum}
\end{figure}

Figure~\ref{fig:spectrum}(a) shows the fluorescence spectra of NV${^-}$ color centers of the ND-doped tellurite glass rod obtained with 532 nm excitation light. The spectra were collected in three geometrical configurations: first, associated with the excitation and fluorescence detection from the rod side (Figure~\ref{fig:spectrum}(c)) corresponding to the spectrum depicted by the blue curve; second, associated with the excitation and detection from the top end of the rod (Figure~\ref{fig:spectrum}(d)) and corresponding to the spectrum marked with the red curve in Figure~\ref{fig:spectrum}(a); third, with the excitation from the side and fluorescence detection from the top presented in Figure~\ref{fig:spectrum}(e) which corresponds to the green line in Figure~\ref{fig:spectrum}(a). Each of these spectra exhibits a 637 nm ZPL peak of NV${^-}$ as shown in detail in inset Figure~\ref{fig:spectrum}(a) which confirms that the developed glass indeed contains nanodiamonds with NV${^-}$ color centers. The peak observed around 555 nm was also present in the undoped glass and only marginally contributes to the overall fluorescence level.  
 
Interestingly, all three spectra of the tellurite rod have their amplitudes in the short-wavelength wing (below 637 nm) significantly higher than the reference spectrum of the used ND sample under the same (532 nm) excitation (Figure~\ref{fig:spectrum}(b)). Since the undoped tellurite glass does not possess any spectral feature in the range 560-800 nm, we have to associate the observed spectral changes with the ND admixture. In addition to the well-evidenced spectrum of NV${^-}$ there are other principally possible sources of the fluorescence below 637 nm: one is the possible green H3 emission, the second is the NV${^0}$ fluorescence with ZPL at 575 nm. The first possibility, however, cannot be realized with the excitation by 532 nm light used in our experiment, thus we assign the enhancement of the short-wavelength fluorescence wing in Figure~\ref{fig:spectrum}(a) to NV${^0}$. The exact nature of this enhancement is an open question at the moment and deserves future detailed studies. Definitely, a thermal conversion between NV${^-}$ and NV${^0}$ and increase of vacancy mobility with temperature play an important role here, as noticed in Ref.~\cite{guo2020adjustable}.
  
For wavelengths above 637 nm, in Figure~\ref{fig:spectrum}(a) different amplitudes of the phonon sidebands can be noticed. They may be caused by the different ratios of NV${^0}$ to NV${^-}$ for various NDs observed in three configurations. The influence of NV${^0}$ to NV${^-}$ shape and intensity of the spectral line is described in a Ref.~\cite{jeske2017stimulated}. The observed spectra may be compared with the fluorescence spectrum of the starting material, MDNVN1 diamond powder, presented in Figure~\ref{fig:spectrum}(b) for various excitation wavelengths. For excitation at 405 nm, we have a characteristic emission peak for green (NVN) color centers at 520 nm; for excitation at 488 nm, the mixed spectra of the green/red color centers are emitting a significant peak at 575 nm associated with NV0 color centers; and the spectrum recorded for 532 nm excitation, has a peak emission at 680 nm characteristic for red fluorescing color centers.

\subsection{ODMR in Magnetic Field and Temperature Measurement}

\begin{figure}
    \centering
    \includegraphics[width=0.80\columnwidth]{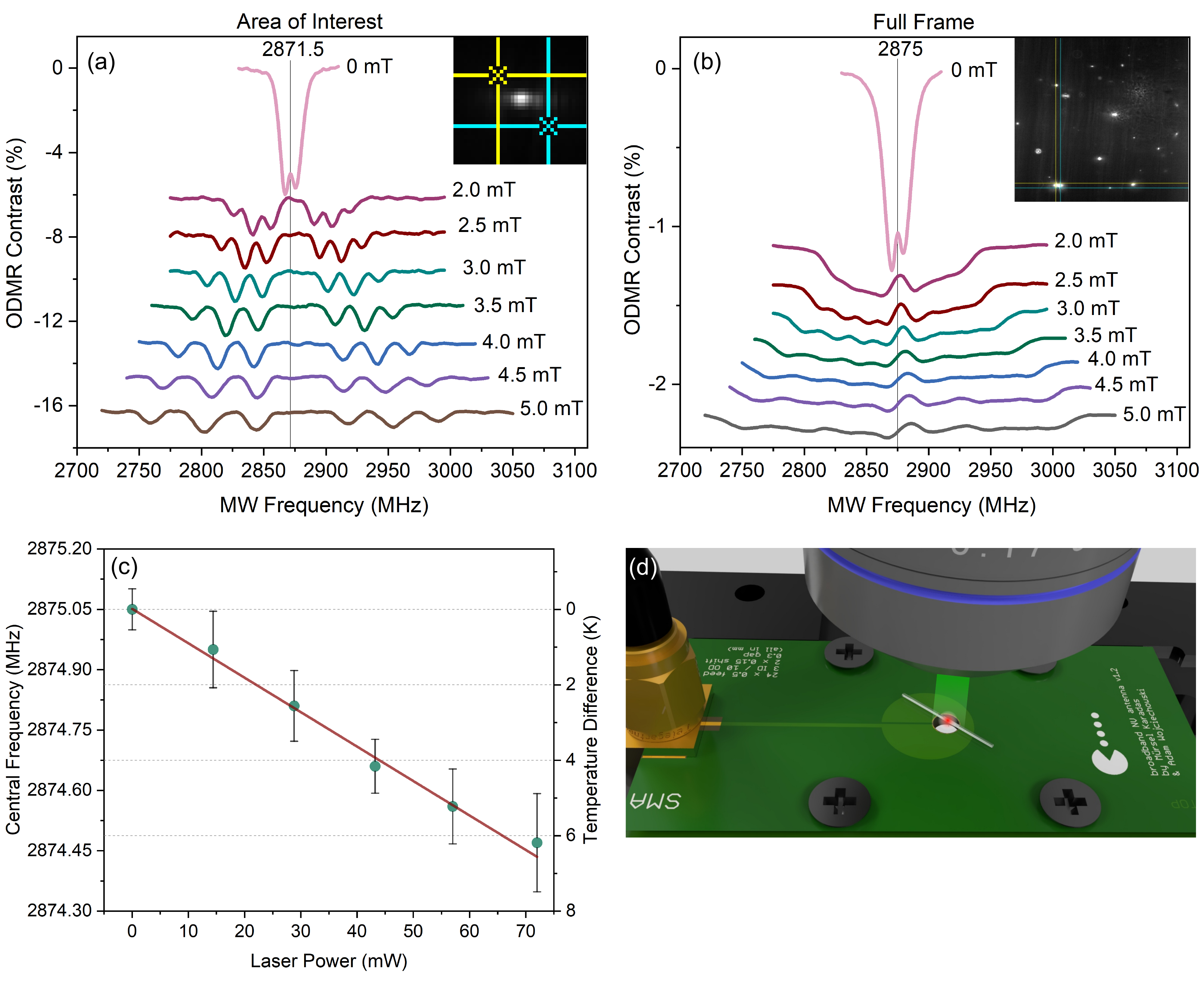}
    \caption{(a) ODMR spectra as a function of the external magnetic field collected from the Area of Interest where a single was particle visible. The inset presents a sensing area of about 24 $\mu$m$^2$ selected with the cursors. (b) ODMR spectra as a function of the external magnetic field collected from the Full Frame area of 205$\times$173 $\mu$m. (c) Dependence of the resonance frequency (left vertical axis) on the heating-laser power (horizontal axis). The temperature difference corresponding to the resonance frequency shift is displayed on the right vertical axis. The straight line shows a linear fit to the collected data. (d) Illustration of the wide-field setup used for ODMR measurements: the tellurite rod was placed on MW antenna and the exciatation and fluorescence light was passing through the same  objective. The illuminated volume contained many NDs.}
    \label{fig:odmr}
\end{figure}

The dependence of the NV energy-level structure on external perturbations enables various sensing applications, with the most important being the magnetic field and temperature probing. The signal detection relies upon shifts of energy levels caused by the magnetic field and temperature, which can be monitored by the corresponding shifts of fluorescence and/or ODMR spectra.

The relevant energies are described by the Hamiltonian of the NV color center which takes into account the interaction of electronic spins with the magnetic field, the thermal expansion of the crystal lattice, and the electron-phonon interactions. The simplified form of the NV Hamiltonian, which ignores the diamond anisotropy and the effects of electric fields has the following form:
\begin{equation}  \label{eq:cfreq}
\ H_{NV} = D S_z^2 + g_{NV} \mu_B \textbf{B} \textbf{S} + \textbf{S} \textbf{A} \textbf{I},\,
\end{equation}
with $D$ as a zero-field splitting parameter along the z-axis equal to 2.87 GHz, $S_z$ being the spin component along the axis of a defect, \(\textbf{S}\) being the electronic spin operator, g the electron Land\'e\ factor, \(\mu_{B}\) the Bohr magneton, \(\textbf{B}\) the applied magnetic field, \(\textbf{A}\) the hyperfine tensor and \(\textbf{I}\) the nitrogen nuclear spin operator. The first and second terms in Eq.\ref{eq:cfreq} represent the main contributions enabling thermometry and magnetometric applications, respectively. 

ODMR measurements of the developed tellurite glass rods containing NV${^-}$ color centers were conducted in a wide-field setup at room temperature. The tellurite rod with nanodiamonds was mounted on a microstrip patch antenna where the MW field around 2.87 GHz was delivered (the measurement configuration is presented in Figure~\ref{fig:odmr}(d)). The MW antenna was designed similarly to the the one described in Ref.~\cite{Sasaki2016}. A movable neodymium magnet was used to create the magnetic field varying in the 0-5.0 mT range and an additional laser (532 nm) was employed to heat the sample for temperature measurements.
 
Figure~\ref{fig:odmr}(a) presents the ODMR spectra collected from the area around 24~$\mu$m$^2$ corresponding to different magnetic field strengths with the individual signals vertically offset for clarity. In this configuration, the maximum ODMR contrast without the magnetic field was 6\% and reduced to around 2\% with the applied magnetic field, consistently with the observation of NV ensembles in bulk diamond mono-crystals~\cite{steinert2010high}. The geometry of the ODMR signal collection in the present work was different from the one reported in Ref.~\cite{ruan2018magnetically} as a large area  was illuminated at once in a wide-field type configuration. This resulted in the detection of the signal from an ensemble of NDs at the same time. Additionally, we have also observed a slightly higher ODMR contrast, which we attribute to an improvement of MW field strength at the probed fiber volume. 

The ODMR signals collected for the camera full-frame area of around 3.5$\times$10$^4\mu$m$^2$, covering dozens of diamond particles, are presented in Figure~\ref{fig:odmr}(b). The observed spectra in a nonzero magnetic field reflect the fact that the sensing volume contains many NV color centers oriented in arbitrary directions relative to the direction of the magnetic field. With a small detection volume, this averaging is not complete and, hence, the ODMR spectra are a mixture of discrete and structure-less components. Despite this orientation-related averaging, the broadened spectra still enable measuring the magnetic field intensity by detecting the outer-most edges of the spectrum~\cite{wojciechowski2019}. 
 
The magnetic sensitivity can be estimated following Ref.~\cite{dreau2011avoiding} as:
\begin{equation}  \label{eq:sens}
  \eta_{DC} = \frac{h\delta}{\sqrt{e/(8ln2)}g_{NV}\mu_{B}C\sqrt{N}},
\end{equation}
where C is the ODMR contrast in a nonzero magnetic field, $C=0.04$, the ODMR linewidth is $\delta= 27\pm0.5$ MHz, and $h$ is the Planck's constant. The number of photons per second detected at the output end of the glass rod was $N=3\times10^6 ~\textrm{s}^{-1}$,  and resulted in a projected magnetic sensitivity of around $10^{-5}$~T~Hz$^{-1/2}$, consistent with the sensitivity obtained for tellurite fibers in Ref.~\cite{ruan2018magnetically}.

Similarly, based on Ref.~\cite{barry2020sensitivity} one can estimate the sensitivity of measuring the temperature of the ND and its environment as:

\begin{equation}  \label{eq:temp1}
  \eta_{T} = \frac{\delta}{|dD/dT|}\frac{1}{C \sqrt{N}},
\end{equation}

where dD/dT is a linear slope, resulting in a projected temperature sensitivity of around 2.2 K/$\sqrt{Hz}$, i.e. consistent with the sensitivity obtained in Ref.~\cite{barry2020sensitivity}. 
 
\section{Conclusion}

Results presented in this work demonstrate that the hybrid material, tellurium glass rods doped with nanodiamonds containing NVN and NV$^{-}$ color centers, is a promising platform for temperature and magnetic field sensing. While the tellurite rods used in this work are larger than previously employed optical fibers, the projected magnetic sensitivity around $10^{-5}$~T~Hz$^{-1/2}$ obtained in this work is comparable to the one achieved in fibers covered with nanodiamonds~\cite{liu2013fiber, fedotov2014fiber} and to that of tellurite fiber incorporated with smaller nanodiamonds studied by Ruan et al.~\cite{ruan2018magnetically}. The bigger size of diamond particles used in this work lead to a significantly larger number of observed fluorescing diamonds that survived the melting and drawing process. We observed the effect of broadening and reducing of the ODMR contrast from 6\% to $\sim$2\% with an increase of the magnetic field, similarly to the results reported in Refs.~\cite{ruan2018magnetically, fedotov2014fiber}. This is caused by the random orientation of individual NDs as well as the increased strain in the diamond lattice due to placement inside the glass matrix. 

The ability to use NVN and NV$^-$ color centers implemented in the tellurite glass increases the versatility of the developed material and enables its wide applications. For example, temperature measurements based on the shift of ZPL of NVN became also possible with the described material as demonstrated in Ref.~\cite{davies1974vibronic}, albeit with a lower sensitivity than the NV$^-$ ODMR measurement enables. For further increasing the sensitivity, the MDNVN1 powder used in this work can be replaced with the one containing NDs with higher concentrations of NV${^-}$ color centers. On the other hand, in light-sensitive applications it may be useful to trade the sensitivity offered by NV$^-$ centers for a blue-shift of optical excitation and detection bands offered by $NVN$ centers.

Temperature measurements based on ODMR studies presented in this paper demonstrate that this type of sensors can be used to simultaneously detect magnetic fields and temperature changes. Spatial dispersion of diamonds inside the tellurite rod can be thus seen as an advantage enabling temperature and magnetic field gradient measurements. Nanodiamonds dispersed in the tellurite glass can thus be a versatile platform for measuring local microscopic magnetic fields and temperature sensing in the microscale. Furthermore, the drawn structure can be easily connected to an undoped optical fibre, offering much better light transmission over larger distances, to act as a remote sensor probe with the ND-doped sensing volume localized only at the fibre end. 

\section{Experimental Section}
\subsection{Sample Preparation}

Nanodiamond incorporated tellurite glass was fabricated using a four-component system with the following chemical composition [mol \%]: TeO$_2$ (73), ZnO (20), Na$_2$O (5), La$_2$O$_3$ (2) (abbreviated as TZNL) identical to that reported in Ref.~\cite{ebendorff2014nanodiamond}. High purity chemical reagents were used to make batches for glass melting. The glass was synthesized at a maximum temperature of 825$^{\circ}$C and stirred twice during melting to homogenize the glass. Next, the glass was cooled down to 650$^{\circ}$C and 2 mg of ND powder was added to 200 g of TZNL, stirred again, and inserted into a furnace at 650$^{\circ}$C for 10 min. The resulted concentration of the diamond powder in the glass was 0.001 $\%$ w/w. Recently, Kumar et al.~\cite{kumar2021high} reported that annealing of NDs at 750 - 1000$^{\circ}$C increases the concentration of NV$^{-}$ centers thanks to the vacancy migration in the lattice. The glass was then cast into a preheated graphite form and annealed from 335$^{\circ}$C to room temperature at a cooling rate of 0.5$^{\circ}$C/min. The drawing process was carried out in a fiber drawing tower at a temperature of 440$^{\circ}$C. Rods about 400 $\mu$m in diameter were obtained and used for ODMR magnetic field and temperature measurements in this study. The ND powder (MDNVN1 from Adamas Nanotechnologies with 750 nm particle size, as determined via DLS measurements) was used as a carrier for H3 and NV$^{-}$ color centers. The use of a powder resulted in random positioning of diamond particles inside the glass matrix and yielded measurable green and red fluorescence from the manufactured tellurite rods. For undoped glasses, the measured refractive index was n$_d$ = 2.0655 and the coefficient of linear thermal expansion measured in a range of 20 - 200$^{\circ}$C was equal to 17.5$\times10^{-6}$~K$^{-1}$. When the undoped tellurite glass was excited with a green light (532 nm) laser, it did not exhibit noticeable fluorescence in the red (600-800 nm) wavelength range as verified with an optical spectrometer (AvaSpec-3648, Avantes). Hence, any fluorescence in the doped glasses is attributed to the introduced NDs.

\subsection{Fluorescence Spectra Collection}

Optical fluorescence spectra presented in ~\ref{fig:spectrum}(a) were recorded using a 0.3 m monochromator (SR303i, Andor - Oxford Instruments) equipped with a 1200 grooves/mm grating and an intensified charge-coupled device (ICCD) detector (DH740, Andor - Oxford Instruments). The spectra in ~\ref{fig:spectrum}(b) were collected with commercial confocal microscope (LSM710, Zeiss) equipped with spectrometer (SP2150, Acton) and ICCD camera (PI-MAX2, Princeton Instruments).

\subsection{Confocal Microscopy and ODMR Technique}

Optical excitation was achieved using a 532 nm laser (Sprout G, Lighthouse Photonics) with an output power of around 1 mW for confocal imaging, and about 60 mW (for wide-field ODMR measurements).
In the home-made confocal setup the oil immersion objective with a magnification of $100x$ (100x/1.3 UPLFLN, Olympus) was used to illuminate and collect fluorescence from the sample. The movement of a sample in three directions with 200x200x200 $\mu$m steps size and resolution of 0.4 nm was enabled by the piezo-scanner stage (Nano-LP200, Mad City Labs). Detection of the emitted fluorescence was performed by a single photon counting module (SPCM-AQRH-14-FC, Excelitas Technologies). The pinhole (30 $\mu$m) placed just before the fibre connected to the SPCM eliminated the fluorescence emitted out of the focal plane by spatial filtering. Instruments of the in-house build confocal setup were controlled by a computer with the Qudi software~\cite{binder2017qudi}. Confocal image $\gamma$ presented in Figure~\ref{fig:confocal}(c) was taken with a commercial wide-field microscope (Axio Observer Z1, Zeiss) equipped with an sCMOS camera (Orca 4.0 V2, Hamamatsu).
The ODMR spectra were collected in a wide-field setup. For wide-field illumination, the green laser was focused onto the back-focal plane of the objective (40x/0.6 LUCPLFLN, Olympus), the optical excitation of NDs was achieved in the same way as with the confocal setup. The collected fluorescence was projected onto a 704x594 pixel photosensitive sensor of a camera (Zyla 5.5 sCMOS, Andor - Oxford Instruments) and the used optics yielded an effective pixel size of about $\sim$0.06 $\mu$m$^2$ with the field of view of 305x244 $\mu$m. The camera was connected to a computer which controlled the experiment and collected data. The MWs were delivered from a signal generator (SG386, Stanford Research Systems).

\medskip
\textbf{} \par 
\medskip
\textbf{Acknowledgements} \par 
The authors would like to acknowledge the financial support of the Foundation for Polish Science under the project Team-NET No. POIR.04.04.00-00-1644/18.

\bibliographystyle{MSP}
\bibliography{bibliography.bib}

\end{document}